\documentclass[11pt]{emulateapj}
\usepackage{natbib}
\usepackage{amsmath,amsfonts,amssymb}
\usepackage{graphicx,pslatex}
\def\idm#1{{\mbox{\scriptsize #1}}}

\newcommand\Ym{\langle Y\rangle}
\newcommand\Chi{{(\chi^2_\nu)^{1/2}}}
\def\astrobj#1{#1\ }

\def\url#1{\texttt{#1}}
\newcommand\pstar{{\astrobj{HD~160691}}}
\newcommand\pstara{{\astrobj{$\mu$~Ara}}}

\begin{document}
%
%
%
\title{Orbital solutions to the HD~160691 ($\mu$~Arae) Doppler signal}
\author{Krzysztof Go\'zdziewski\altaffilmark{1},
        Maciej Konacki\altaffilmark{2}, and
        Andrzej J. Maciejewski \altaffilmark{3}
       }
\altaffiltext{1}{Toru\'n Centre for Astronomy,
  N.~Copernicus University,
  Gagarina 11, 87-100 Toru\'n, Poland; k.gozdziewski@astri.uni.torun.pl}
\altaffiltext{1}{California Institute of Technology,
  Pasadena; maciej@gps.caltech.edu}
\altaffiltext{3}{Institute of Astronomy,
  University of Zielona G\'ora,
  Podg\'orna 50, 65-246 Zielona G\'ora, Poland;
  maciejka@astro.ia.uz.zgora.pl}

\begin{abstract}
We perform a dynamical analysis of the recently updated set of  the radial
velocity (RV) measurements of the HD~160691 ($\mu$~Arae). The  purely
kinematic, 2-Keplerian model of the measurements leads to the best-fit 
solution in which the eccentricity of the outer planet is about 0.7 and its
semi-major axis is about 4~AU. The parameters of the inner planet are  well
determined. The eccentricity is about~0.3 and the semi-major axis is about
1.65~AU. In such circumstances, the best 2-Keplerian model leads to a 
catastrophically unstable configuration, which is self disrupting in less than
20,000~yr. In order to derive dynamically stable configurations which are 
simultaneously consistent with the RV data, we use the so called GAMP (Genetic 
Algorithm with MEGNO Penalty) which incorporates the genetic fitting
algorithms  and a verification of the stability by the fast indicator of the
dynamics  (MEGNO). Using this method, we derive meaningful limits on the
parameters of  the outer planet which also provide a stable behavior of the
system. It appears,  that the best-fit solutions are located in a shallow
valley of $\Chi$, in the  $(a_{\idm{c}},e_{\idm{c}})$-plane,  extending over
2~AU (for the formal $1\sigma$ confidence interval of the best fit). We find
two equally good best-fit solutions leading to the qualitatively  different
orbital configurations.  One of them corresponds to the center of the 5:1 mean
motion resonance (MMR) and the second one describes a configuration  between
the 6:1 and 17:2~MMRs.  Generally, the \pstar system can be found in  the zone
confined to other low-order MMRs of the type $p$:1 with $p>5$.   Our results
support the classification of the \pstara as a hierarchical   planetary system,
dynamically similar to other known multi-planet systems  around HD~12661 and
HD~169830. The results of the GAMP analysis on the extended  data set are fully
consistent with our previous conclusions concerning a much  shorter
observational window. It constitutes a valuable confirmation that the applied
method is well suited for the analysis of the RV data series  which only
partially cover the longest orbital period. 
\end{abstract}
%
%
\section{Introduction}
%
%
Precise measurements of the radial velocity (RV) of two stars, HD~160691
(\pstara) and HD~154857, from the Anglo-Australian survey for extrasolar
planetary systems have been recently published by~\cite{McCarthy2004}. Our
particular attention is devoted to the \pstar system.  \cite{Butler2001}
reported a planet with a 2~yr orbital period about the \pstara.  This discovery
was confirmed by \cite{Jones2002b} and the observational window extended to
4~yr revealed an additional linear trend in the RV signal, interpreted as the
signature  of a more distant planet. The data presented in the new paper by the
Anglo-Australian  Planet Search team cover about 6~yr. This extended
observational window allows  to confirm the existence of the second Jovian
planet in the system.  Very recently,  \cite{Santos2004} discovered a small,
$14~$-Earth masses object in a circular orbit,  at the distance of only 0.09~AU
from the \pstara star and confirmed the findings of \citep{McCarthy2004}. The
Anglo-Australian team determined the  Keplerian elements of the Jovian
companions: the orbital periods of 645~d, 3000~d  and the eccentricities of
0.2, 0.57, respectively \citep{McCarthy2004}.  According to these authors, the
current data set covers about 70\% of the  longer period.

The detection of multiple, Jovian-like planets in stages from shorter to longer
orbital periods are obviously expected from the Doppler surveys and
\citep{McCarthy2004} announce another possible multiple system around
HD~154857. It is interesting to follow the subsequent estimations of the
orbital parameters of such system, derived on the basis of an increasing number
of observations. In the preprint by \cite{Jones2002a}, the preliminary orbital
solution for \pstara suggested a proximity of the two putative planets to the
2:1~mean motion resonance (MMR).  Such configuration would be the third
occurrence of this particular low-order resonance among only about 15
multi-planet systems known at that time. However, the formal 2-Kepler solution
found  in the preprint appeared to be very unstable. In the published paper,
\cite{Jones2002b} excluded the possibility of the 2:1~MMR due to an increased
orbital period of the outer planet (by about 300~d). Our independent analysis
of the same RV data set \citep{Gozdziewski2003e} has revealed that in the
specific case of \pstara, when one has only a part of the orbital RV signal of
the additional planetary companion, the application of 2-Kepler  model of RV
signal certainly leads to artifacts. The formally best-fit configurations are
self-disrupting on a short time-scale! The newest set of measurements leads to
the 2-Kepler solution with the outer planets' period increased to $\simeq
3000$~d~\citep{McCarthy2004}. This 2-Keplerian fit {\rm still} describes a
configuration which disintegrates after only $20,000$~yr.  It can be easily
checked by direct integration of the 3-body problem. In fact even a simple
analysis shows that this best fit is located  in the proximity of the planetary
collision zone. The exact collision curve  can be determined through  
$a_{\idm{b}} ( 1 + e_{\idm{b}} ) = a_{\idm{c}} ( 1 - e_{\idm{c}} )$.  This
means that the large eccentricity of the companion~c would be possible only if
an orbital dynamical mechanism protecting the planets from an encounter
existed. Another possibility is that the real orbital elements are in fact 
different from those determined by \cite{McCarthy2004}.

Clearly, one needs a method of dynamical analysis of the orbital fits which
will  reproduce an observed RV signal and simultaneously fulfill stability
criteria. In our method, we not only account for the mutual interactions
between the planets~\citep{Laughlin2001} but also eliminate the orbital fits
corresponding to strongly chaotic orbits, regardless of their goodness of fit
expressed by $\Chi$. We treat the dynamical behavior in terms of the chaotic
and regular (or weakly-chaotic) states as an observable, at the same level of
importance as the RV measurements are. Details of the algorithm which will be
called from hereafter the Genetic Algorithm with MEGNO Penalty (GAMP) are
presented in \citep{Gozdziewski2003e}. This method seems to be particularly
useful for systems undergoing strong mutual interactions such as the \pstara
system. According to the quoted preliminary estimates of the semi-major axes of
the companions and their eccentricities, the likely configurations are related
to the zone of strong low-order mean motion resonances like 5:2, 4:1, 5:1, 6:1,
7:1. The widths of these resonances rapidly increase with the growing
eccentricity of the outer companion, finally  leading to their overlapping and
a creation of a zone of global instability (see section~4 in this work). In
this instance, the MMRs-originating chaotic diffusion leads to violent,
macroscopic changes in the physical state of the system \citep{Froeschle1999}. 
In a planetary system, it means close encounters between the planets and/or the
parent star, collisions or ejection of a body from the system. Following this
reasoning, to find configurations  which are stable over long periods of time,
it is essential to eliminate the strongly chaotic solutions. These solutions
originate in the unstable MMRs and lead to  a discontinuity of the searched
parameter space, in the sense of the dynamical stability. The sophisticated
structure of the phase space  can be understood in terms of the KAM theorem.
Obviously, both the kinematic models of the Doppler signal as well as the
gradient-like methods of the RV data fitting which require the $\Chi$ to be a
continuous function of its variables, are not well suited to this case. On the
contrary, the discontinuous structure of the orbital parameter space justifies
the application of non-gradient fitting and Monte-Carlo based techniques. 

Taking into account these very basic considerations, we analyzed the RV data
published by \cite{Jones2002b}. We found that the quasi-linear trend present in
the reflex motion of \pstar permitted a continuum of equally good solutions in
the $(a_{\idm{c}},e_{\idm{c}})$-plane. Moreover, using the GAMP algorithm  and
assuming that $a_{\idm{c}}<5$~AU,  it was possible to reduce the space of
stable fits, permitted by the RV data, to a number of solutions confined to the
zone of low-order mean motion resonances (MMRs) like 3:1, 7:2, 4:1, 9:2, 5:1,
11:2, 6:1. Moreover, the short observational window did not permit a
distinction between them. At present, having the access to the updated
observations, we have an occasion to verify these early predictions and to
test  the reliability of the fits obtained with GAMP. Many stars accompanied by
Jovian planets reveal linear trends in the Doppler signal indicating the
presence of additional companions~\cite[e.g.,][]{Butler2003}. Hence, we have a
chance to test if our method is helpful in obtaining preliminary and realistic
estimates of their orbital parameters. 

\section{The numerical setup}
As the basic tool for searching for the best fit solutions to the RV data, we
use the genetic algorithm scheme (GA) implemented by \cite{Charbonneau1995}  in
his publicly available code named 
PIKAIA\footnote{\url{http://www.hao.ucar.edu/public/research/si/pikaia/pikaia.html}}.
The synthetic reflex motion of the star is described by the 2-Keplerian RV
model~\citep[as in][]{Gozdziewski2003e} and the self-consistent Newtonian,
$N$-body  model \citep{Laughlin2001} incorporating the MEGNO indicator
penalty~\citep{Gozdziewski2003e}. It is known that the genetic scheme is not 
efficient for finding very accurate best-fit solutions but provides good
starting fits for much faster and more precise gradient methods, like the
Levenberg-Marquardt  (LM) algorithm~\citep{Stepinski2000}.  Nevertheless, the
GA makes it possible, in principle, to find the global minimum of $\Chi$ while
the gradient methods are basically  local. In our experiments, the GA fits were
finally refined by a yet another very accurate non-gradient minimization scheme
by Melder and Mead \citep{Press1992}, widely known as the simplex method.  The
fractional convergence tolerance to be achieved in the simplex code has been
set in the range $10^{-6}$--$10^{-9}$. Typically, the lower accuracy has been
forced by time consuming GAMP tests. The application of the simplex refinement
in the CPU expensive GAMP code is essential because it reduces the CPU usage
dramatically (by a factor of ten, at least).

For the stability analysis, we use two numerical tools which are helpful in
resolving the regular (quasi-periodic) and irregular (chaotic) character of a
tested configuration. The already mentioned MEGNO
algorithm~\citep{Cincotta2000,Cincotta2003,Cincotta2004} is already used in a
few our papers. In this work (also to avoid a routine approach in our stability
studies), we apply an alternative method derived on the basis of the Fast
Fourier Transform (FFT). This spectral method (SM) was developed by
\cite{Michtchenko2001}. Its idea is very simple: in the quasi-periodic system,
the spectral signal, (here $f(t) = a(t) \exp(\mbox{i}\lambda(t)$ where $a(t)$
and $\lambda(t)$ are the heliocentric canonical elements \citep{Laskar1995},
the semi-major axis and mean longitude, respectively), has only a small number
of leading frequencies while in the chaotic system the spectrum of these
frequencies becomes very complicated. By counting the number of the frequencies
(the spectral number, $SN$) in the FFT spectrum of the signal that have
amplitudes over a given noise level (typically, we use a threshold of about a
few percent of the largest amplitude), one can distinguish between a regular
and chaotic motion. It appears that these two fast indicators are similarly
well sensitive to the chaotic diffusion generated by the MMRs in the systems
with Jupiter-like planets and the required integration time is relatively very
short, typically, about $10^4$ periods of the outermost planet.  The spectral
method seems to have many advantages. Its  simplicity is appealing. Under
certain conditions, the SM is even more efficient than MEGNO because one avoids
integrating complex variational equations.  It provides a straightforward
identification of the MMRs.  To summarize the differences between the two
methods, one could say that the SM detects the chaotic diffusion as an
"analogue" device, while the MEGNO (Lyapunov exponent type indicator) is more
of a "digital" tool. Details of the comparison of these two complementary
methods will be described elsewhere.

\section{Searching for the Newtonian best fits}

First, we performed the analysis of the 2-Keplerian fits to the RV data of the
\pstar system similarly to  \citep{Gozdziewski2003e}. The best fit solution has
$\Chi=1.644$ and the rms of $4.147$~m/s (see Table~1). This fit is
catastrophically unstable. It is not surprising because  $e_{\idm{c}} \simeq
0.72$ by far exceeds eccentricities corresponding to the collision curve for
the two planets. Our 2-Keplerian fit has significantly larger $e_{\idm{c}}$ and
the orbital period of the outer planet than found by the discovery team. At
present, $P_{\idm{c}}\simeq 3360$~d is also longer than from the previous
2-Keplerian estimate \citep{Jones2002b,Gozdziewski2003e}. This conclusion and
our experiments (not only from this work) seem to follow an empirical rule:  if
the measurements cover only a fragment of the longest orbital period, widening
the observational window leads to an even longer period. Obviously, the current
set of RV data still does not permit to identify a stable solution, when
assuming the purely kinematic 2-Keplerian model of the RV data. The synthetic
RV curve of the best 2-Keplerian  fit in Fig.~\ref{fig:fig1} is barely
different from the curve corresponding to the synthetic curve derived from the
$N$-body model (more details will be given further). Yet, they describe
completely different configurations. Obviously, their $\Chi$  and rms must be
almost the same. This example clearly shows that the smallest $\Chi$ or rms
cannot be the decisive factor for the choice of the proper description of the
observed planetary configuration.

In the next experiments, we used only the full $N$-body model of the reflex
motion of the parent star. The first question is how (if at all) the
self-consistent fits  (for the moment---without the instability penalty) change
the character of the best fit solutions.  Such fits have been initialized by
the GA and refined by the simplex. The results are illustrated in
Fig.~\ref{fig:fig2}.  It shows the projections of the best fits to which the
simplex converged in every individual run (expressed by  osculating
heliocentric elements) on the planes of the relevant orbital parameters. 
Hundreds of  independent runs of the fitting code have been preformed. The
osculating elements are given at the moment of the first observation. The best
fit solution is given in Table~\ref{tab:tab1} as the NL1 fit. Its rms is about
4.15~m/s and $\Chi\simeq 1.643$, thus almost the same as for the 2-Keplerian
model. Yet, the derived minimal mass of the outer planet is much smaller than
in the 2-Keplerian solution. 

The collected fits make it possible to obtain a general view of the orbital 
configurations of the system permissible by the $N$-body model.  It appears
that the elements of the inner planet are already well fixed. This is not the
case for the outer companion. Its eccentricity, in the range of $1\sigma$
confidence interval of the best fit solution NL1, can vary over 0.1-0.8 and the
semi-major axis spreads over 2-3~AU. Also  the mass of the outer planet cannot
be precisely estimated. Surprisingly, the angular elements of both companions
are relatively well bounded.  Assuming a coplanar configuration, this result
makes it possible to reduce  the space of barely determined parameters to three
dimensions: $(M_{\idm{c}},a_{\idm{c}},e_{\idm{c}})$. This is quite comfortable
for the studies of the  system dynamics because it can be limited to a
representative $(a_{\idm{c}},e_{\idm{c}})$-plane parameterized by the mass of
the outer companion. The assumption of coplanarity may be justified by the
recent results \citep{Santos2004}. As we already mentioned, these authors
discovered a small, Neptune-like planet around the \pstar.  They also measured
the projection of the  rotational velocity, $v\sin i$, of \pstar and found it
close to the theoretical, maximal value. Let us suppose that the orbital plane
of the smallest planet is almost perpendicular to the rotational axis of
\pstar. This means that the orbit is almost "edge-on". In the presence of a
possibly highly inclined Jovian planet, the orbit of the smaller companion
would be  influenced by the Kozai resonance. For a  large relative inclination,
this mechanism  forces a large eccentricity of the inner, smaller body and
finally can destabilize its motion. Because the eccentricity of the inner
planet is very small~\citep{Santos2004},  the two most inner orbits should not
be significantly inclined.   Assuming that the planetary system emerged from a
coplanar proto-planetary disk,  the whole  system is likely close to the
coplanar configuration. Still, the recent results by \cite{Thommes2003}  and
\cite{Adams2003} indicate that a large fraction of planetary systems which
interact gravitationally can be in fact significantly non-coplanar.
Unfortunately, the currently available data do not permit to investigate such a
possibility.

As we have already noticed,  the NL1 solution is formally as good as the fit
obtained  with the 2-Kepler model.  Unfortunately, there is also a similar
problem with the stability of the relevant configuration. The evolution of the
orbital elements (Fig.~\ref{fig:fig3}) clearly shows that the motion is
strongly chaotic and, in fact, corresponds to collisional orbits. In the
specific integration,  the Lyapunov time is only 233~yr. It will be shown later
that this fit corresponds to the border of 5:1~MMR (Fig.~\ref{fig:fig10}).

In order to estimate the formal errors of the best fit NL1, we followed 
\cite{Gozdziewski2003d}. The actual error of the determination of the RV
consists of two parts: the measurement error and the contribution from  the
variability of the star itself. For chromospherically quiet G and K dwarfs the
stallar jitter is about the level of the measurements errors. In this paper we
adopted $\sigma_{\idm{jitter}} \simeq 3$~m/s, following the estimates of the
stellar variability of G dwarfs by \cite{McCarthy2004} (see their Fig.~1).
These estimates are based on 6~yr monitoring of a few representative G-type
stars similar to \pstara. To estimate the error of the best fit NL1, we added
to every original measurement a random Gaussian noise with the zero mean value
and the mean dispersion $\sigma \simeq 4$~m/s and the Gaussian noise of the
stellar jitter with  $\sigma_{\idm{jitter}} \simeq 3$~m/s. Next, the best fit
was recalculated with  the simplex algorithm using the synthetic data set. This
procedure was repeated 5000~times. After finding the mean values of the best
fit parameters, we calculated their dispersions. These dispersions are used as
the estimates of the mean errors of the osculating elements and shown in
parentheses in Table~1. These estimates confirm our earlier findings indicating
that the planetary phases can be relatively well fixed while the mostly
unconstrained elements are the mass, the semi-major axis and the eccentricity
of the outer planet. 

Further, to see whether the formal RV errors increased by a contribution from
the stellar jitter influence the overall distribution of the best fit elements,
we repeated the GA search using the original data points and their mean errors
changed to  $\sigma \rightarrow  \sqrt{\sigma^2 + \sigma_{\idm{jitter}}^2}$. 
The parameters of the best fit found in this search are given in Table~1 as the
NL2 solution and the distribution of the best fit elements is shown in
Fig.~\ref{fig:fig4} and Fig.~\ref{fig:fig5}. Obviously, the larger mean errors
do not lead to a significant modification of the overall distribution of the
best fits. Let us note that the projections shown in
Fig.~\ref{fig:fig2},\ref{fig:fig4} and \ref{fig:fig5} serve as an alternative,
realistic estimation of the fit errors. The errors obtained in this way are
even larger than the previously derived. 

The experiments utilizing the $N$-body model of the RV measurements confirm
that the existence of the second planetary companion. Nevertheless, all the
best-fit initial conditions still lead to self-disrupting configurations and,
additionally, are not well constrained.   Recalling the discussion in the
introduction and the results of dynamical analysis in \cite{Gozdziewski2003e},
we already  have a rough picture of the phase space in the relevant zone of
$(a_{\idm{c}},e_{\idm{c}})$.  This zone is confined to strong, unstable
low-order MMRs. Due to the preferably large initial $e_{\idm{c}}$, the chances
of finding  stable configurations by a straightforward Keplerian or Newtonian
fit to the RV data are poor.

Thus, to find the best-fit orbital solution in a zone of stability, we have
made extensive calculations by applying the GAMP. The code has been restarted
hundreds of times. The control parameters of the code have been changed and
adjusted to give an additional degree of randomness in the GA search. The
results are illustrated in Fig.~\ref{fig:fig6}.  As in the previous cases, this
figure shows a projection of the found best-fit solutions onto different 
planes of osculating elements. Only  those fits are shown which have $\Chi <
1.668$. This limit corresponds to the formal  $1\sigma$ confidence interval of
the best fit solution (GM1, see Table~1) having $\Chi=1.64701$ and  rms
1.147~m/s. As we can expect, the stable solutions have $e_{\idm{c}}<0.5$. This
strongly confirms our prediction from \cite{Gozdziewski2003e}.

The reader may notice that the $\Chi$ for GM1 (Table~1) is given with an
unusually high precision of 5 decimal places. This is due to another solution
(GM2, see Table~1) having almost the same $\Chi$ while its $a_{\idm{c}}$ is
shifted by 0.4~AU. The orbital behavior of the corresponding configurations
(see Fig.~\ref{fig:fig7} and Fig.~\ref{fig:fig8}) clearly shows that they
belong to qualitatively different dynamical regimes.  The GM1 fit corresponds
to a very stable, regular behavior which is confirmed by the MEGNO rapidly
converging over 1~Myr. Surprisingly, the GM2 fit describes a system whose
stability is marginal. This system is trapped into the 5:1~MMR. This is
illustrated in panel Fig.~\ref{fig:fig7}c showing one of the critical arguments
of the resonance, 
$\sigma_{\idm{5:1}}=
5\lambda_{\idm{c}}-\lambda_{\idm{b}}-\varpi_{\idm{c}}-3\varpi_{\idm{c}}$.

\section{Dynamics and stability analysis} 
 
In principle, the best-fit solutions found by the GAMP search should be
dynamically stable. The chaos in the GM2 solution can be explained by a
relatively short  integration time used in the algorithm (about
1000~$P_{\idm{c}}$ at 5~AU) which does not allow to eliminate mildly chaotic
configurations. We assume here that the chaos manifestating through a linear
growth of MEGNO after a relatively long time (much longer than adopted in the
GAMP) is regarded as a week one. This can be considered as a desired property
of the method rather than its drawback. In fact, it can hardly be expected that
multi-planet extrasolar systems found so far should be strictly quasi-periodic
and regular, in terms of the Lyapunov exponent. Nevertheless, due to the fast
convergence of the MEGNO and its great sensitivity to the chaotic evolution,
the short integration times make it possible to eliminate the strongly chaotic
configurations which very likely would lead to significant changes of the
orbital elements. To verify the long-term stability of the best fit solutions
found in this way, we should always analyze the structure of the phase space in
their vicinity. Such analysis (using numerical or analytical tools) helps us to
explore the dynamical  environment of the nominal configuration, estimate
whether the fits are robust  to the errors and the found configurations do not
change qualitatively under  small adjustments of the initial conditions. 

First, we have examinated the GM1 fit by computing the stability map in the
$(a_{\idm{c}},e_{\idm{c}}$)-plane using the SM code. The relevant orbital
parameters were varied and other elements were fixed at the values quoted in
Table~1. The results of this test are illustrated in Fig.~\ref{fig:fig9}. The
integration time for this map is $\simeq 10^5$~yr ($\simeq 10^4$ orbital
periods at 5~AU). The best fit solution is marked by a big circle.  The
planetary collision line for fixed initial $a_{\idm{b}},e_{\idm{b}}$ is 
depicted as a smooth curve. The dynamics over this line or in its proximity is
very chaotic and the planets' eccentricities grow rapidly to $1$, indicating
collisions or ejections from the system. As we could expect, the stability map
reveals an  amazingly complex and beautiful structure of MMRs. These MMRs can
be easily identified. It appears that the GM1 fit is close to the border of the
11:2 and 17:2~MMRs. In the previous paper on the \pstar system, we already
suggested that this system may be very similar do the HD~12661 planetary
system. It appears that, according to the current best-fit solutions, the
dynamical environment of \pstar may be a close analogue of this system. Let us
recall that in \cite{Gozdziewski2003d} we found that the HD~12661 may be
located close to the 6:1~MMR while the orbital fits of \cite{Butler2003} are
related to the proximity of 11:2~MMR \citep{Gozdziewski2003c}. Due to the
mentioned tendency that  the estimated orbital period of the outermost body
increases with a longer  observational time span, the proximity of the \pstar
system to 6:1~MMR may  also be quite possible.

A similar stability test has been performed for the GM2 solution
(Fig.~\ref{fig:fig10}).  The upper panel is for the SM stability map in the
($a_{\idm{c}},e_{\idm{c}})$-plane. The middle panel is for the maximum
eccentricity of the planet~c attained during the integration period. In both
plots the best fit is marked with a big circle.  For comparison, in the
stability map (Fig.~\ref{fig:fig10}, upper panel),  the solution corresponding
to NL1 fit is marked with a smaller circle. Note that due to very similar
orbital elements in the NL1 and GM2 fits, depicting both of them at the same
stability map is reasonable. Now it is clear why the NL1 fit leads to a chaotic
configuration. According to the identification of MMRs in this map, the GM2 fit
is inside the libration island of the 5:1~MMR, while the NL1 fit lies in a
close vicinity or within the separatrix of this resonance.

In the $\max e_{\idm{c}}$-map (Fig.~\ref{fig:fig10}, the middle panel), we also
marked (small circles) all fits gathered in the GAMP search within $1\sigma$
confidence interval of the best fit solutions GM1, GM2.  Such illustration of
the best fit solutions is justified because, as we already mentioned,
$(a_{\idm{c}},e_{\idm{c}})$-plane is representative for the dynamical state of
the system. The distribution of the best fits in this plane makes it possible
to derive some interesting conclusions. In spite of the fact that the points
representing the fits are spread over $a_{\idm{c}} \in [4,6)$~AU, the tendency
of a decreasing $e_{\idm{c}}$ with larger values of $a_{\idm{c}}$ is clear.
Extrapolating this trend, the outer planet should have a small  initial
eccentricity for large $a_{\idm{c}}\simeq 6$--$7$~AU. The acceptable fit 
values within the $1\sigma$ error of $e_{\idm{c}}$ can be varied over 0--0.4.  
Similarly to the previously analyzed LN1 and LN2 fits, the distribution of  the
osculating elements in the ($a_{\idm{c}},e_{\idm{c}}$)-plane gives us  an
insight into realistic error estimates of these orbital elements.

Obviously, all the relevant fits "avoid" the neighborhood of the collision
zone. Close to and over the collision line,  $\max e_{\idm{c}}$ grows to 1
which means that the system should be disrupted. The comparison of the
stability maps for the GM1 and GM2 shows that the width of the MMRs
significantly depends on the particular orbital parameters. Nevertheless, their
positions do not change much. Let us recall, that the stability maps in
Fig.~\ref{fig:fig9} and \ref{fig:fig10} correspond to $a_{\idm{c}}$ altered by
about 0.5~AU and $e_{\idm{c}}$ by $\simeq 0.1$. 

The last panel shown in Fig.~\ref{fig:fig10} refers to the maximum of the
argument $\theta=\varpi_{\idm{c}}-\varpi_{\idm{b}}$ attained during the
integration time. Librations of $\theta$ correspond to the so called secular
apsidal resonance (SAR) when the planetary apsides remain aligned or
anti-aligned.  The behavior of $\theta$ is important for the discussion of the
dynamics of 2-planet configurations in the absence of strong MMRs. In this
instance, the motion of the planetary system may be averaged over the mean
longitudes. The secular behavior of the averaged system can be described in
terms of  the secular octupole theory of hierarchical planetary systems by
\cite{Lee2003} and the recent more general secular theory of 2-planet systems
by \cite{Michtchenko2004}. According to the results of these authors, the
generic 2-planet coplanar system which is far from low-order MMRs and collision
zones exhibits a long-term stable behavior. This behavior can be characterized
by the time evolution of $\theta$. This angle can librate about $0^{\circ}$ or
$180^{\circ}$, circulate, or librate about $0^{\circ}$ in the
large-eccentricity regime corresponding to the true secular resonance 
\cite[for details see][]{Michtchenko2004}. When the eccentricities are
moderate, the two first possibilities (modes) are alternative for the secular
evolution of the planetary system. Under some conditions~\citep{Lee2003}, the
SAR regime appears with a probability close to~1.

The octupole secular theory by \cite{Lee2003} is well suited for the \pstar
system in the range of $a_{\idm{c}} > 4$~AU and under the assumption that a
given initial condition (IC) is distant from low-order MMRs. In this case,  
the best-fit solutions are characterized by a relatively small ratio 
$\alpha=a_{\idm{b}}/a_{\idm{c}} \simeq 0.3$, as for hierarchical
configurations. The bottom panel in Fig.~\ref{fig:fig10}  illustrates the
extent of the SAR in the range of $a_{\idm{c}} \in [4,6]$~AU. Besides the
centers of the MMRs, the SAR with  the anti-aligned apsides appears over an
extended range of $a_{\idm{c}}$ and for relatively small $e_{\idm{c}}$.  Let us
recall that this map has been derived numerically, simultaneously with
calculating the SM stability map. It is very well reproduced by means of the
octupole theory, see the map in the left panel in Fig.~\ref{fig:fig11}, which
is obtained by the numerical integration of the averaged equations of motion
\citep{Lee2003,Lee2003a}. Another  SAR map of the secular system, computed in
the eccentricities plane, is shown in the right panel of Fig.~\ref{fig:fig11}.
In both of these maps, the best fit GM1 is found in the extended zone of
circulations of $\theta$. Figure~\ref{fig:fig12} illustrates the evolution of
$\theta(e_{\idm{b}})$ and $\theta(e_{\idm{c}})$ for the IC defined by GM1. 
These plots have been derived by the numerical integration of the full $N$-body
equations of motion. The phase plots are computed for the constant values of 
the semi-major axes and the total angular momentum, $J$.  Let us recall that
the averaged system is governed by 1~degree of freedom Hamiltonian in which
$\theta$ and  the eccentricity of one planet are the relevant canonical
variables. To construct the phase plots, we fixed the initial $\theta$ either
at $0^{\circ}$ or $180^{\circ}$ and $e_{\idm{b}}$ was varied appropriately
while the other initial eccentricity, $e_{\idm{c}}$,  was derived from the $J$
integral. The fact that the phase curves are not ideally smooth can be
explained by the proximity to the two MMRs, 11:2 and~17:3. This confirms  that
the GM1 fit corresponds to the configuration with $\theta$ confined to the
extended zone of circulation. The large amplitudes of eccentricities shown in
Fig.~\ref{fig:fig8} can be also explained by the  phase plots.

The secular features of the  GM1 configuration remind us of another
hierarchical system about HD~169830. The dynamical environments of their
secular configurations are very similar. For instance, compare the discussed
figures with the corresponding ones from \cite{Gozdziewski2004} (their Fig.~14
and Fig.~15). The appearance of the SAR in extrasolar systems with well
separated giant companions is not an extraordinary feature as sometimes
claimed. In the light of the mentioned secular theories, in the regime of
moderate eccentricities, the secular systems should be found in one of two
distinct regimes of stable motion characterized by the circulation of
librations of the critical argument~$\theta$. Nevertheless, one should be aware
that the secular theories give us only an approximate description of the
long-term behavior. In the proximity to the strong MMRs, the stability should
be investigated by an application of theories suitable for every individual
resonance or long-term direct integrations. 

\section{Conclusions}

The new RV data of the \pstar system published by the Anglo-Australian Planet 
Search team \cite{McCarthy2004} enable us to refine the orbital best-fit
solutions. The preliminary observations published in~\cite{Jones2002b}
suggested the presence of the second companion inducing a linear trend in the
RV signal. In  the light of the currently available data, it is very likely
that the semi-major axis of the outer planet cannot be smaller than about
$4$~AU. This makes the system relatively separated. Curiously, the kinematic
2-Keplerian model of the  RV signal constantly produces artifacts e.g. the
apparent proximity to the 2:1~MMR  or large eccentricities of the outer
companion. These configurations lead to  collisions between the planets in a
very short time. In this work, we applied  the "dynamical" fitting by forcing a
requirement that a real system should not be disrupted over a short time-span
(thousands of years). Our analysis with the proposed GAMP algorithm reveals
that the current data set still does not well constrain the semi-major axis of
the outer companion. But the requirement of dynamical stability puts strong
limits on the initial eccentricity of the outermost planet. The current set of
observations is equally well modeled by a continuum of statistically equivalent
best-fit solution which are distributed over $a_{\idm{c}} \in [4,6)$~AU and
$e_{\idm{c}}$ about $[0-0.4]$, along a flat valley of $\Chi$  in the
$(a_{\idm{c}},e_{\idm{c}})$-plane of osculating orbital parameters. 
Surprisingly, the initial longitudes of the coplanar configurations seem to be
already well fixed.

The dynamical analysis of the best-fit solutions reveals that the currently
determined parameters of the outer planet are confined to the zone of low order
MMRs with the inner planet, like 5:1, 11:2, 6:1, 13:2, 7:1. The distribution of
the best solutions, within the formal $1\sigma$ confidence interval of the best
fit, reveals a clear correlation between initial values of $a_{\idm{c}}$ and
$e_{\idm{c}}$---the larger semi-major axis of the outer planet, the smaller its
initial  eccentricity. According to the empirically observed law, the increase
of the observational window results in the increase of the determined orbital 
periods. This would mean that in the current best fit $a_{\idm{c}}\simeq
4.79$~AU  is still smaller than the real value. It is difficult to predict  the
real $a_{\idm{c}}$ but very likely it is in the range $[5,6]$~AU.

During the final preparation of this paper, we have learned about a discovery 
of a Neptunian-mass body in the \pstar system, close to the parent star 
\citep{Santos2004}. The semi-major axis of this small planet is about
$0.09$~AU.  Obviously, it is interesting to verify whether the presence of this
body can affect the best-fit solutions and the dynamical behavior of the whole
planetary system. The RV measurements published by the Anglo-Australian Planet
Search team have the mean errors comparable to the RV semi-amplitude variations
imposed by the new planet (4~m/s). For the moment, the newest observations, as
reported  in the discovery paper, span a very short period of time (at present,
about 80~d). Unfortunately, these measurements are not available.  They would
be very helpful in improving our fits because they cover the end-part of the
longer period and should constrain the parameters of the outer companion. We
tried to refine our fits using the data from a scanned figure published by the
Swiss team on their
WWW~page\footnote{\url{http://obswww.unige.ch/~udry/planet/hd160691.html}}. 
The full data set, produced by a combination of  the Anglo-Australian
measurements and these synthetic RV points, has been firstly modeled using the
$N$-body signal. We used velocity offsets different for the two observatories.
Still, we encountered similar problems as in the case of the Anglo-Australian
data itself. The best $N$-body fits for the 3-planet  system converge to
$a_{\idm{c}}\simeq 3.8$~AU and $e_{\idm{c}}> 0.6$. According to the stability
analysis, such fits locate the Jovian planets in a strongly chaotic zone. For
the 3-planet best-fit solutions, we have $\Chi \simeq 1.56$ and the rms of
$\simeq 3.6$~m/s. Assuming that the innermost planet does not  disturb the
motion of the Jovian companions too much and applying the GAMP for  these two
planets only (otherwise, the CPU time requirement is very significant),  we
derived solutions qualitatively similar to the GM1 and GM2 fits (although  the
rms of these solutions exceeds $4.7$~m/s). They tend to have a large
$e_{\idm{c}} \simeq 0.5$. Simultaneously, through the requirement of stability,
the GAMP places the best fits about $a_{\idm{c}} \simeq 5.5~AU$. Thus, in spite
of a better time coverage, the curious inconsistency between the pure  $N$-body
and the GAMP-derived solutions is still evident. These results can  only be
preliminary. A reliable modeling of the RV data is a difficult task  without
having access to the actual measurements.

Although the elements of the innermost planet seem to be determined, due to the
barely determined parameters of the outer planet, we think that  the dynamical
studies concerning the whole system do not make at present a great sense.  As
we have shown, the two Jovian planets can be coupled by a strong, low-order
MMRs or can be well separated and relatively far from MMRs.  These situations
create completely different dynamical environments for the whole planetary
system. 

In terms of the overall dynamical behavior, the \pstara system appears to be
similar to the HD~12661 and to the HD~169830. The Jovian planets in these
systems are well separated. The variation of eccentricities spans a similar,
relatively extended range from 0 to 0.5-0.6. Their semi-major axes ratio
corresponds to a proximity of low-order MMRs, like 6:1 or 11:2 (for the
HD~12661) or 7:1, 8:1 (similar to the HD~169830). Whether a real link exists
between these systems is not clear at the moment. We still have to wait for new
observations which eventually make it possible to derive significantly accurate
determination of their parameters. 

Using the GAMP algorithm we performed a very preliminary  analysis of the RV
data of the HD~154857 system. Similarity to the \pstar, these data reveal the
presence of one Jovian planet  and a linear trend, indicating the presence of
the second, more distant body. Our $N$-body fits  make it possible to
approximate the elements of this planet: its mass is about 0.5~M$_{\idm{J}}$,
semi-major axis about $4$~AU and large eccentricity $\simeq 0.78$. Of course,
these elements are barely constrained. Nevertheless, the parameters of the
inner planet are quite well fixed and we can already compute the stability map
of this system.  It is shown in Fig.~\ref{fig:fig13}. The relevant zone is
filled up by strong MMRs, similarly to the \pstar system.  By the same
reasoning, as applied to the \pstar system, the eccentricity of the outer
planet should be strongly limited by the requirement of dynamical stability. 

Our work advocates a dynamical analysis of the RV data. Even if the
observational window  covers only partially the longest orbital period, we are
able to derive a lot of characteristics of the  discussed system, avoiding
falsifying the orbital elements by the purely kinematic approach. In fact,  the
analysis presented in this paper shows that the application of the GAMP-like
algorithm is essential for finding reliable solutions which well reproduce the
RV data and simultaneously lead to orbitally stable configurations of the
planetary companions.

\section{Acknowledgments}
This work is supported by the Polish
Committee for Scientific Research,  Grant No.~2P03D~001~22. K.G. is also
supported by the UMK grant 428-A. M.K. is also supported by NASA through grant
NNG04GM62G.
\bibliographystyle{apj}
\bibliography{ms}

\begin{thebibliography}{25}
\expandafter\ifx\csname natexlab\endcsname\relax\def\natexlab#1{#1}\fi

\bibitem[{{Adams} \& {Laughlin}(2003)}]{Adams2003}
{Adams}, F.~C. \& {Laughlin}, G. 2003, Icarus, 163, 290

\bibitem[{{Butler} {et~al.}(2003){Butler}, {Marcy}, {Vogt}, {Fischer}, {Henry},
  {Laughlin}, \& {Wright}}]{Butler2003}
{Butler}, R.~P., {Marcy}, G.~W., {Vogt}, S.~S., {Fischer}, D.~A., {Henry},
  G.~W., {Laughlin}, G., \& {Wright}, J.~T. 2003, \apj, 582, 455

\bibitem[{{Butler} {et~al.}(2001){Butler}, {Tinney}, {Marcy}, {Jones}, {Penny},
  \& {Apps}}]{Butler2001}
{Butler}, R.~P., {Tinney}, C.~G., {Marcy}, G.~W., {Jones}, H.~R.~A., {Penny},
  A.~J., \& {Apps}, K. 2001, \apj, 555, 410

\bibitem[{{Charbonneau}(1995)}]{Charbonneau1995}
{Charbonneau}, P. 1995, \apjs, 101, 309

\bibitem[{{Cincotta} {et~al.}(2003){Cincotta}, {Giordano}, \& {Sim\'o,
  C.}}]{Cincotta2003}
{Cincotta}, P.~M., {Giordano}, C.~M., \& {Sim\'o, C.} 2003, Physica D, 182, 151

\bibitem[{{Cincotta} \& {Sim{\' o}}(2000)}]{Cincotta2000}
{Cincotta}, P.~M. \& {Sim{\' o}}, C. 2000, \aaps, 147, 205

\bibitem[{{Froeschl{\' e}} \& {Lega}(1999)}]{Froeschle1999}
{Froeschl{\' e}}, C. \& {Lega}, E. 1999, in NATO ASIC Proc. 522: The Dynamics
  of Small Bodies in the Solar System, A Major Key to Solar System Studies,
  463--+

\bibitem[{{Giordano} \& {Cincotta}(2004)}]{Cincotta2004}
{Giordano}, C.~M. \& {Cincotta}, P.~M. 2004, \aap, 423, 745

\bibitem[{{Go{\' z}dziewski} \& {Konacki}(2004)}]{Gozdziewski2004}
{Go{\' z}dziewski}, K. \& {Konacki}, M. 2004, \apj, 610, 1093

\bibitem[{{Go{\' z}dziewski} {et~al.}(2003){Go{\' z}dziewski}, {Konacki}, \&
  {Maciejewski}}]{Gozdziewski2003e}
{Go{\' z}dziewski}, K., {Konacki}, M., \& {Maciejewski}, A.~J. 2003, \apj, 594

\bibitem[{{Go{\' z}dziewski} \& {Maciejewski}(2003)}]{Gozdziewski2003d}
{Go{\' z}dziewski}, K. \& {Maciejewski}, A.~J. 2003, \apjl, 586, L153

\bibitem[{{G}o{\' z}dziewski(2003)}]{Gozdziewski2003c}
{G}o{\' z}dziewski, K. 2003, \aap, 398, 1151

\bibitem[{{Jones} {et~al.}(2002{\natexlab{a}}){Jones}, {Butler}, {Marcy},
  {Tinney}, {Penny}, C., \& B.}]{Jones2002a}
{Jones}, H., {Butler}, P., {Marcy}, G., {Tinney}, C., {Penny}, A., C., M., \&
  B., C. 2002{\natexlab{a}}, MNRAS, astro-ph/0206216

\bibitem[{{Jones} {et~al.}(2002{\natexlab{b}}){Jones}, {Paul Butler}, {Marcy},
  {Tinney}, {Penny}, {McCarthy}, \& {Carter}}]{Jones2002b}
{Jones}, H.~R.~A., {Paul Butler}, R., {Marcy}, G.~W., {Tinney}, C.~G., {Penny},
  A.~J., {McCarthy}, C., \& {Carter}, B.~D. 2002{\natexlab{b}}, \mnras, 337,
  1170

\bibitem[{{Laskar} \& {Robutel}(1995)}]{Laskar1995}
{Laskar}, J. \& {Robutel}, P. 1995, Celestial Mechanics and Dynamical
  Astronomy, 62, 193

\bibitem[{{Laughlin} \& {Chambers}(2001)}]{Laughlin2001}
{Laughlin}, G. \& {Chambers}, J.~E. 2001, ApJ, 551, L109

\bibitem[{{Lee} \& {Peale}(2003{\natexlab{a}})}]{Lee2003a}
{Lee}, M.~H. \& {Peale}, S.~J. 2003{\natexlab{a}}, \apj, 597, 644

\bibitem[{{Lee} \& {Peale}(2003{\natexlab{b}})}]{Lee2003}
---. 2003{\natexlab{b}}, \apj, 592, 1201

\bibitem[{McCarthy {et~al.}(2004)}]{McCarthy2004}
McCarthy, C. {et~al.} 2004, \apj

\bibitem[{{Michtchenko} \& {Ferraz-Mello}(2001)}]{Michtchenko2001}
{Michtchenko}, T. \& {Ferraz-Mello}, S. 2001, ApJ, 122, 474

\bibitem[{{Michtchenko} \& {Malhotra}(2004)}]{Michtchenko2004}
{Michtchenko}, T.~A. \& {Malhotra}, R. 2004, Icarus, 168, 237

\bibitem[{{Press} {et~al.}(1992){Press}, {Teukolsky}, Vetterling, \&
  {Flannery}}]{Press1992}
{Press}, W.~H., {Teukolsky}, S.~A., Vetterling, W.~T., \& {Flannery}, B.~P.
  1992, Numerical Recipes in C. The Art of Scientific Computing (Cambridge
  Univ. Press)

\bibitem[{Santos {et~al.}(2004)}]{Santos2004}
Santos, N.~C. {et~al.} 2004, \aap, 426, L19

\bibitem[{{Stepinski} {et~al.}(2000){Stepinski}, {Malhotra}, \&
  {Black}}]{Stepinski2000}
{Stepinski}, T.~F., {Malhotra}, R., \& {Black}, D.~C. 2000, ApJ, 545, 1044

\bibitem[{{Thommes} \& {Lissauer}(2003)}]{Thommes2003}
{Thommes}, E.~W. \& {Lissauer}, J.~J. 2003, \apj, 597, 566

\end{thebibliography}
%
%
\begin{table}
\tabletypesize{\scriptsize}
\caption{
Initial conditions  of the \pstar system. The first column is for the best
2-Kepler fit expressed in Jacobi elements,  according to the model in
\cite{Gozdziewski2003e}. Fit marked by NL1 is derived by the self-consistent
$N$-body fitting and expressed by osculating heliocentric Keplerian elements at
the moment of first observation (JD=2,450,915.2911). The NL2 solution is
derived by the same approach, when the measurement errors are scaled by  the
stellar jitter ($\simeq 3$~m/s). Fits denoted by GM1 and GM2 are formally the
best fits obtained in the GAMP search. The osculating heliocentric Keplerian
elements are given at the date of first observation. The mass of the parent
star is equal to $1.08~\mbox{M}_{\sun}$.
}
\smallskip
\begin{tabular}{lcccccccccc}
\hline
&\multicolumn{2}{c}{2K} &
 \multicolumn{2}{c}{NL1} &
 \multicolumn{2}{c}{NL2} &
 \multicolumn{2}{c}{GM1} &
 \multicolumn{2}{c}{GM2} \\
Parameter \hspace{1em}
& \ \  {\bf b} \ \  & \ \ {\bf c} \ \ 
& \ \  {\bf b} \ \  & \ \ {\bf c} \ \   
& \ \  {\bf b} \ \  & \ \ {\bf c} \ \   
& \ \  {\bf b} \ \  & \ \ {\bf c} \ \   
& \ \  {\bf b} \ \  & \ \ {\bf c} \ \   \\
\hline
$\mbox{m}_2 \sin i$ [M$_{\idm{J}}$] \dotfill 
				&  1.675  &  7.675    
				&  1.674 (0.068)  &  2.707 (0.898)
				&  1.683  &  5.689  
				&  1.670  &  2.981 
				&  1.677  &  2.398 
\\
a [AU] \dotfill 		&  1.499  &  4.551   
				&  1.497 (0.007)  &  4.325 (0.560)
				&  1.499  &  4.683
				&  1.497  &  4.788 
				&  1.498  &  4.364 
\\
P [d] \dotfill 			& 644.7 &  3368.7
				&         & 
				&         &  
				&         &  
				&         &         \\
e \dotfill     			& 0.202   & 0.719  
				& 0.201 (0.029)   & 0.470 (0.215)   
				& 0.203   & 0.585 
				& 0.204   & 0.340  
				& 0.201   & 0.411      
\\
$\omega$ [deg]\dotfill 		& 115.14  & 348.53  
				& 114.83 (6.37)  & 341.14 (16.23) 
				& 115.15  & 347.03 
				& 116.89  & 343.15
				& 116.42  & 342.70  
\\
$T_{\idm{p}}$ [JD-2,450,000]  \dotfill
				& 2196.8 & 3681.4 
				&         &   
				&         &   
				&         &       \\
$M$ [deg] \dotfill 		& 	  &  
				& 4.68 (4.54)    & 49.79 (26.13) 
				& 3.67    & 66.50
				& 1.45    & 65.2
				& 2.75    & 46.42 
\\
$\Chi$  \dotfill 		& \multicolumn{2}{c}{1.64378} & 
		 		  \multicolumn{2}{c}{1.64464} &
		 		  \multicolumn{2}{c}{1.12871} &
		 		  \multicolumn{2}{c}{1.64701} &
		 		  \multicolumn{2}{c}{1.64707} 
\\
RMS [m/s] \dotfill 	& 	  \multicolumn{2}{c}{ 4.15}   & 
				  \multicolumn{2}{c}{ 4.15}   &
				  \multicolumn{2}{c}{ 4.12}   &
				  \multicolumn{2}{c}{ 4.17}   &
				  \multicolumn{2}{c}{ 4.15}   
\\
$V_0$ [m/s] \dotfill 	& 	  \multicolumn{2}{c}{-37.2}   &
				  \multicolumn{2}{c}{-22.08 (6.12)}  &
				  \multicolumn{2}{c}{-33.37}  &
				  \multicolumn{2}{c}{-22.08}  &
				  \multicolumn{2}{c}{-16.49}  
\\ 
\hline
\end{tabular}
\label{tab:tab1}
\end{table}

%
%

%
\setcounter{figure}{0}
%
%
\begin{figure}[th]
\centering
\includegraphics[width=14cm]{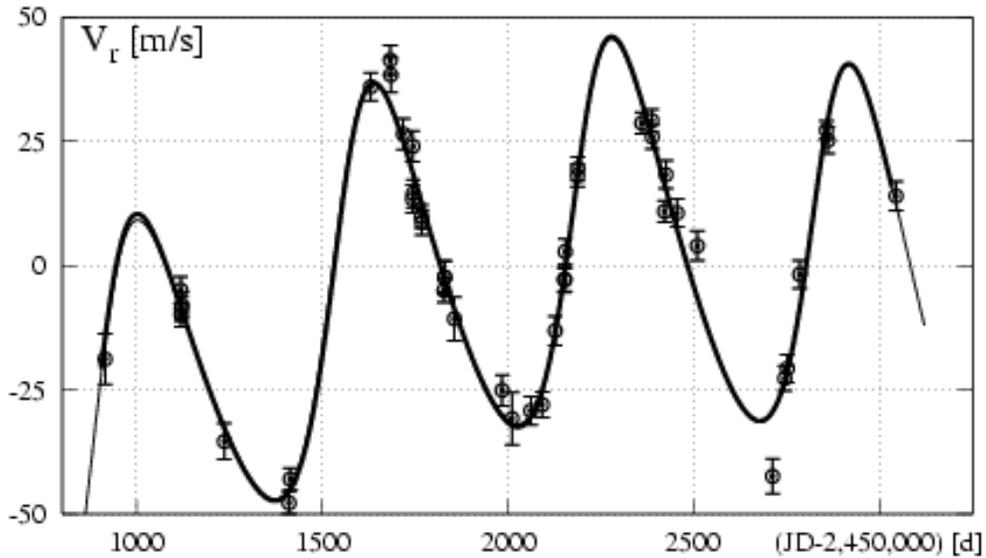}
\caption{
Synthetic radial velocity curves for the \pstar system. The thin line is for
the best 2-Keplerian solution (2K, see Table~1).  The thick line is for a
stable, $N$-body fit~GM1 (see the text for explanation). Open circles are for
the RV measurements published in \citep{McCarthy2004}. }
\label{fig:fig1}
\end{figure}

\eject
%
%
\begin{figure}[th]
\centering
\includegraphics[width=15cm]{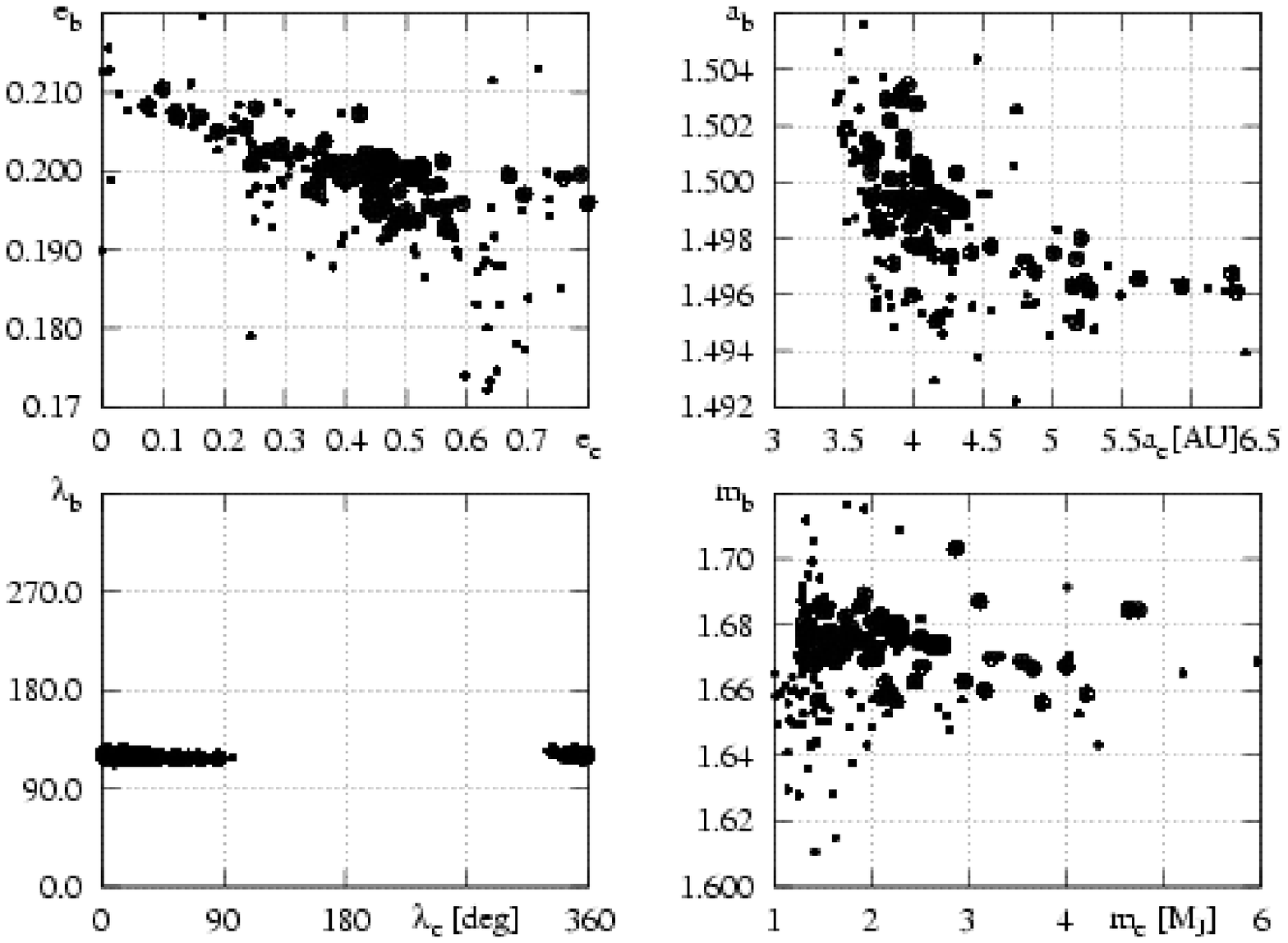}
\caption{
The best fits obtained by the Newtonian model for the RV data published in
\cite{McCarthy2004}. Small filled circles are for solutions with $\Chi$ within
the formal $3\sigma$ confidence interval of  the NL1 fit (see Table~1,
$\Chi<1.75$). Bigger, open circles are for  $\Chi<1.667$ corresponding to
$1\sigma$ confidence interval of the best fit~NL1. Largest filled circles are
for the fits of  $\Chi<1.646$, marginally larger, by about 1\%, from the $\Chi
\simeq 1.644$ of the best fit solution~NL1.
}
\label{fig:fig2}
\end{figure}
%
%
\begin{figure}
\centering
\includegraphics[]{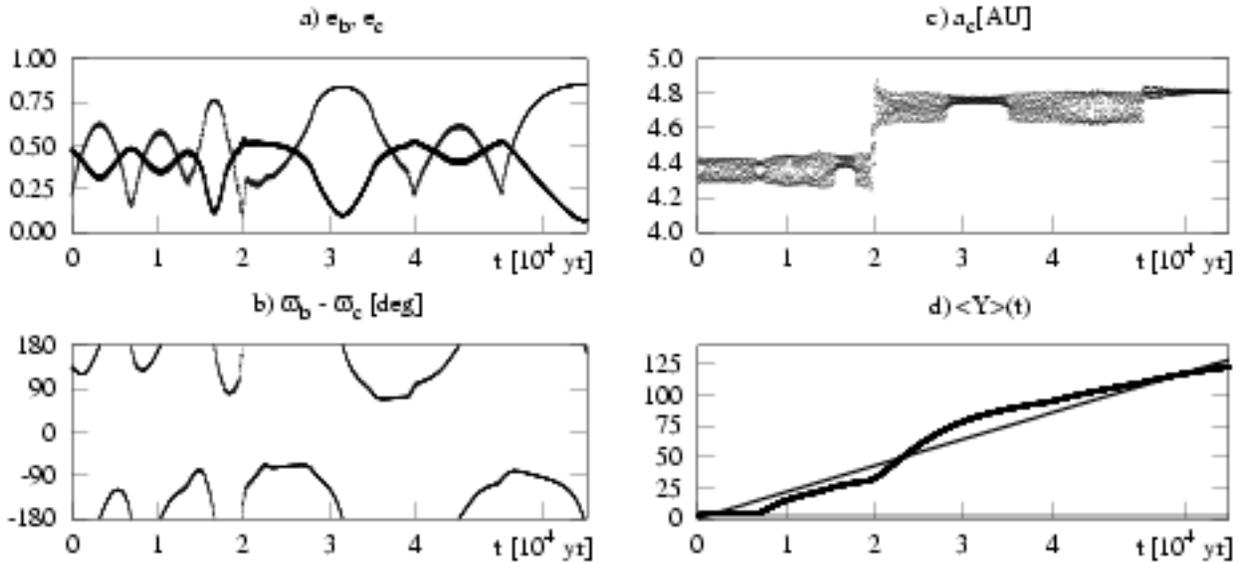}
\caption{
Orbital behavior in the best self-consistent, Newtonian fit NL1.
Panel~a) is for the eccentricities.
Panel b) is for the argument of the SAR.
Panel c) is for the semi-major axis of the outer companion.
Panel d) is for the MEGNO, $\Ym$.
The thin line illustrates the linear fit
$\Ym(t)=(\lambda_{\idm{max}}/2)t+d$.
The derived Lyapunov time, $T_L=1/\lambda_{\idm{max}}$, is about 233~yr.
}
\label{fig:fig3}
\end{figure}
\eject
%
%

\begin{figure}
\centering
\includegraphics[width=15cm]{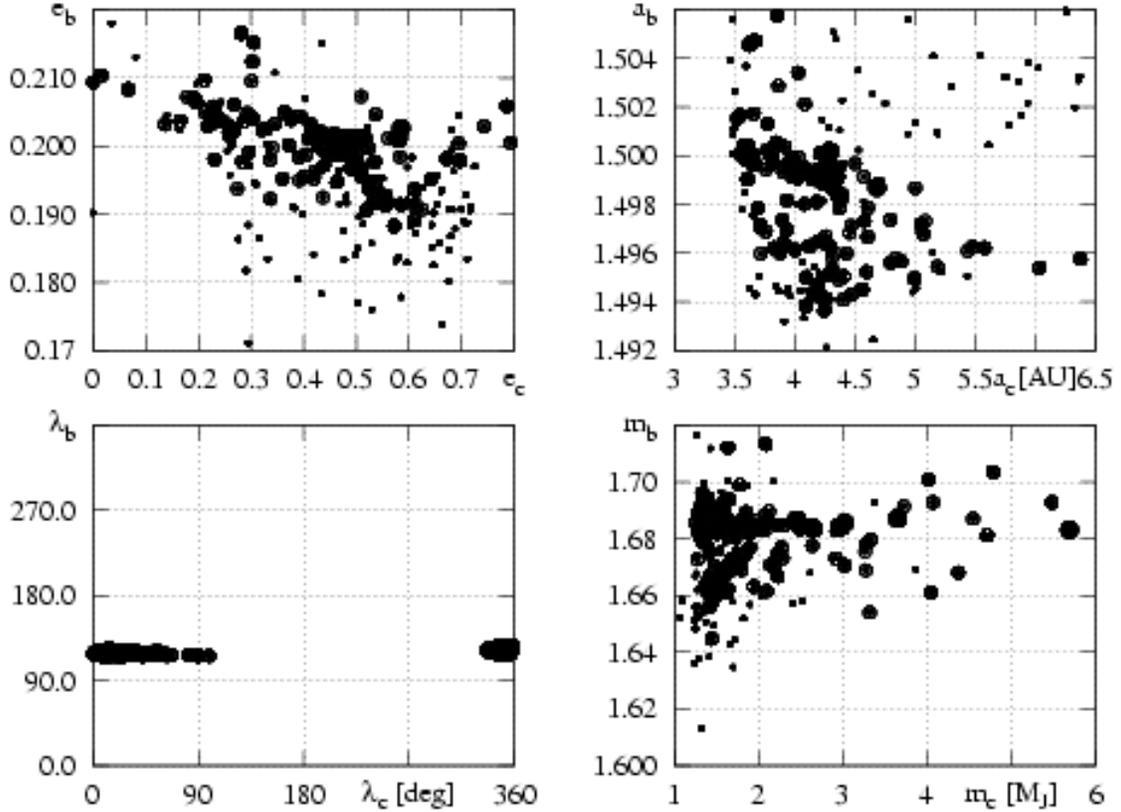}
\caption{
The best fits obtained by the Newtonian model of the RV data and their errors
scaled by the stellar jitter with the mean dispersion about 3~m/s  (see the
text for explanation).  Small filled circles are for solutions with
$\Chi<1.277$, within the formal $3\sigma$ confidence interval of the NL2 fit
(see Table~1). Bigger, open circles are for  $\Chi< 1.16$ corresponding to
$1\sigma$. Largest filled circles are for the fits of  $\Chi<1.130$, marginally
larger, by about 1\%, from the $\Chi \simeq 1.129$ of NL2.
}
\label{fig:fig4}
\end{figure}
%
%
\begin{figure}
\centering
\includegraphics[width=15cm]{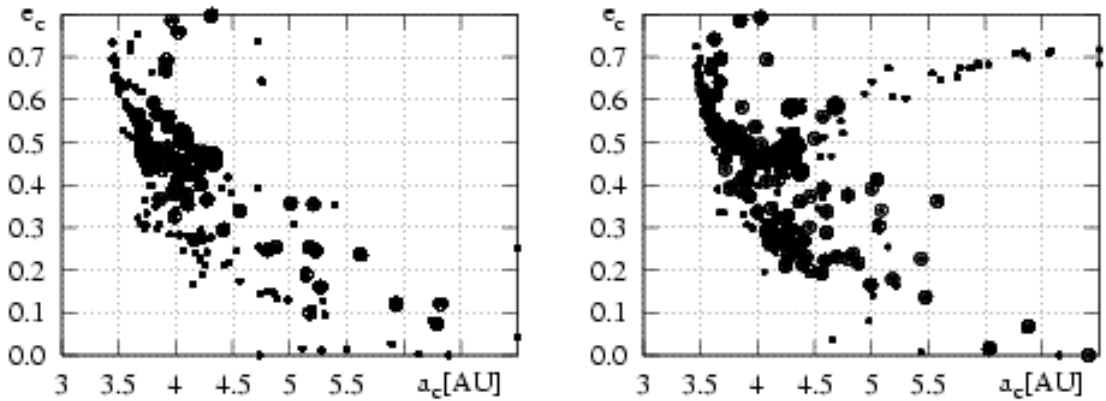}
\caption{
A comparison of the distribution of the best fits  obtained by the $N$-body
fitting in the $(a_{\idm{c}},e_{\idm{c}})$-plane.  The left panel is for the
original data set, the right panel is for the case when the  measurements error
are scaled by the stellar jitter (see the text for explanation). The meaning of
symbols is the same as in Fig.~\ref{fig:fig2} and~\ref{fig:fig4}, respectively.
}
\label{fig:fig5}
\end{figure}
\eject
%
%
\begin{figure}
\centering
\includegraphics[width=15cm]{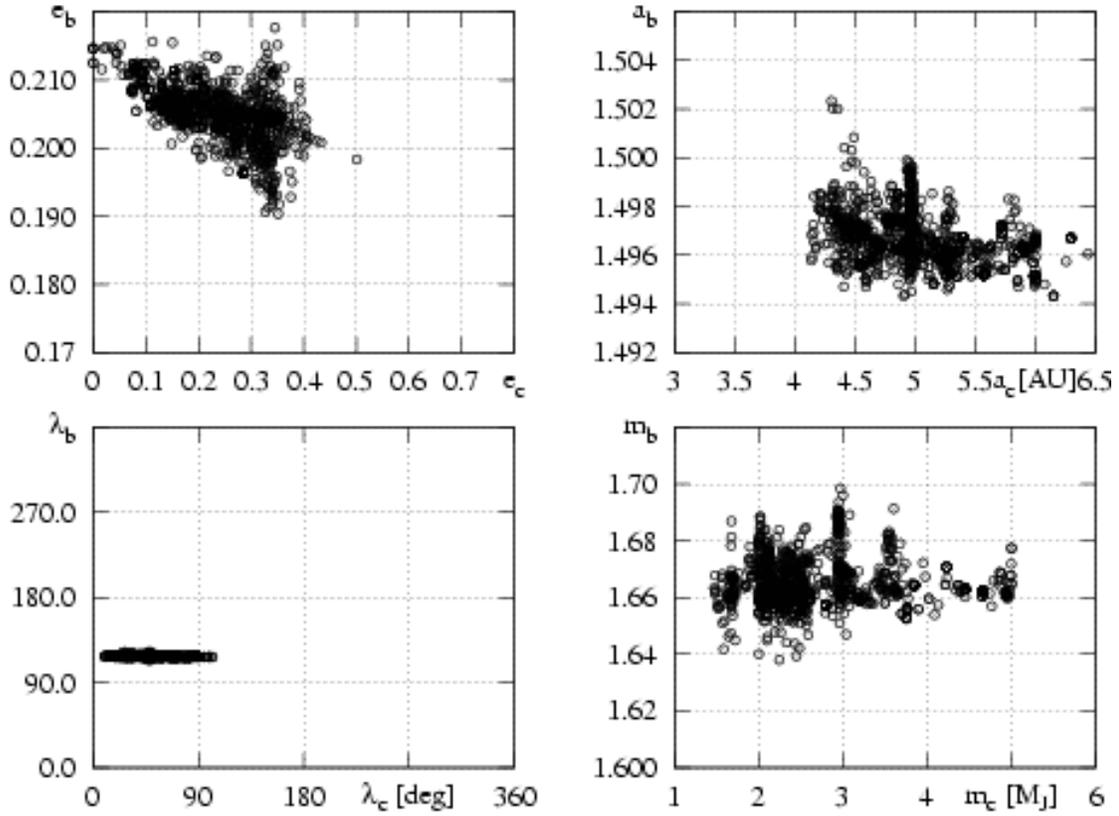}
\caption{
The best fits obtained by the GAMP.  Open circles are for  $\Chi<1.668$, i.e.,
within formal $1\sigma$ confidence interval of the best fit solution having
$\Chi=1.647$.
}
\label{fig:fig6}
\end{figure}
%
%
\begin{figure}
\centering
\includegraphics[]{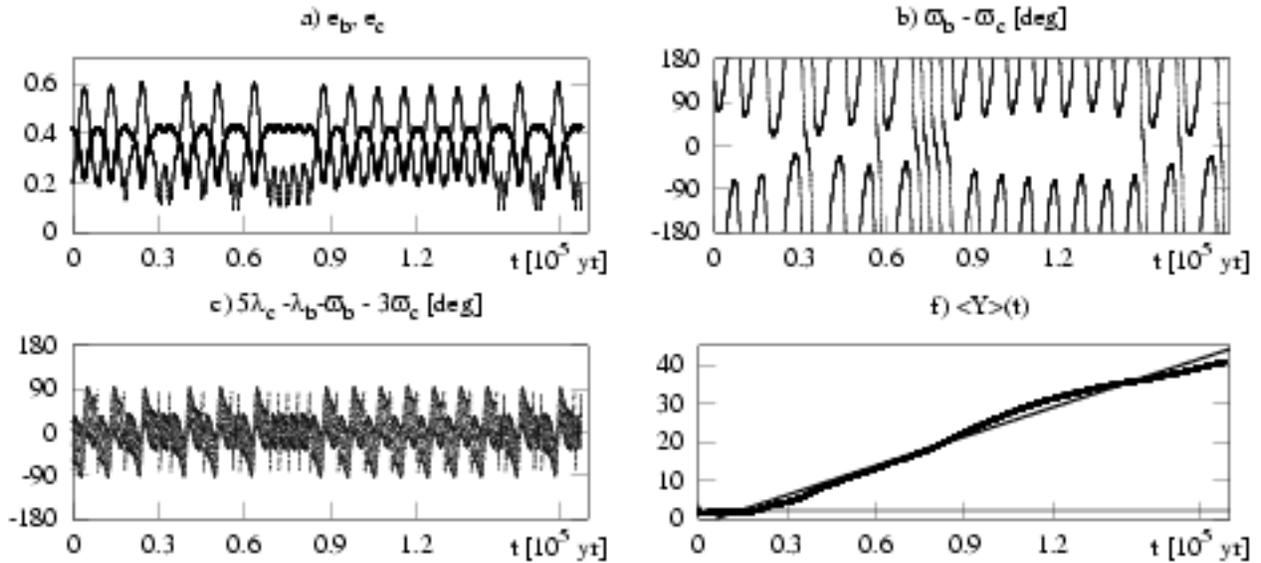}
\caption{
Orbital behavior in the GM2 fit.
Panel~a) is for the eccentricities.
Panel b) is for the argument of the SAR.
Panel c) is for the critical argument of the 5:1~MMR.
Panel d) is for the MEGNO, $\Ym$.
The thin line illustrates the linear fit
$\Ym(t)=(\lambda_{\idm{max}}/2)t+d$.
The derived Lyapunov time,  $T_L=1/\lambda_{\idm{max}}$, is about 35,000~yr.
}
\label{fig:fig7}
\end{figure}
%
%
\begin{figure}
\centering
\includegraphics[]{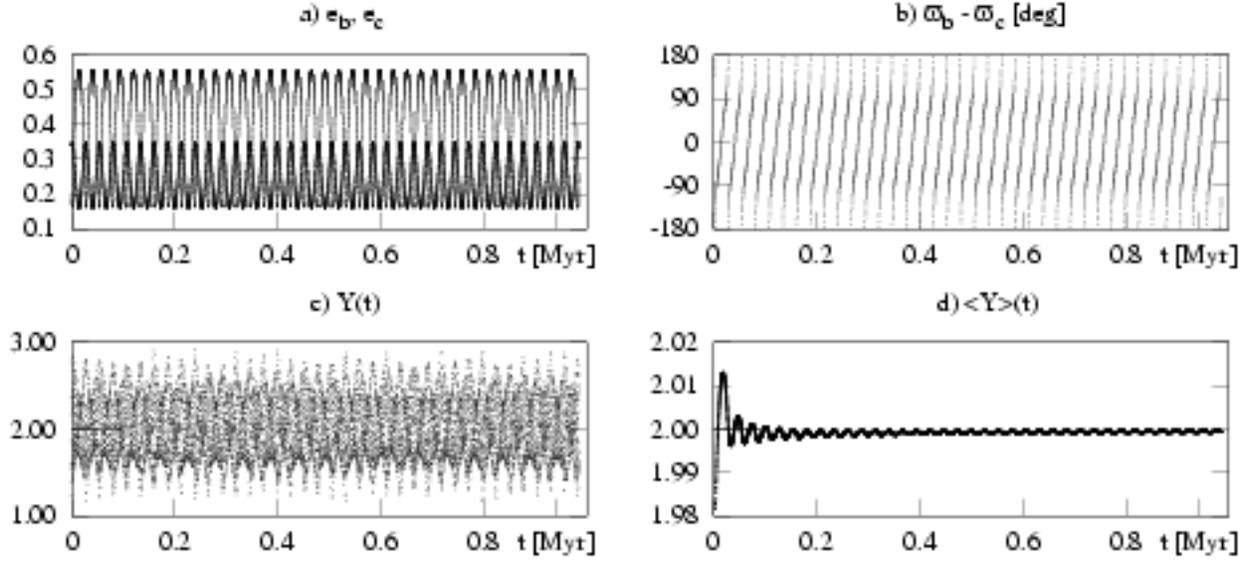}
\caption{
Orbital behavior in the GM1 fit.
Panel~a) is for the eccentricities.
Panel b) is for the argument of the SAR.
Panel c) is for the the temporal MEGNO, $Y(t)$.
Panel d) is for the MEGNO, $\Ym$.
}
\label{fig:fig8}
\end{figure}
%
%
\begin{figure}
\centering
\includegraphics[width=17cm]{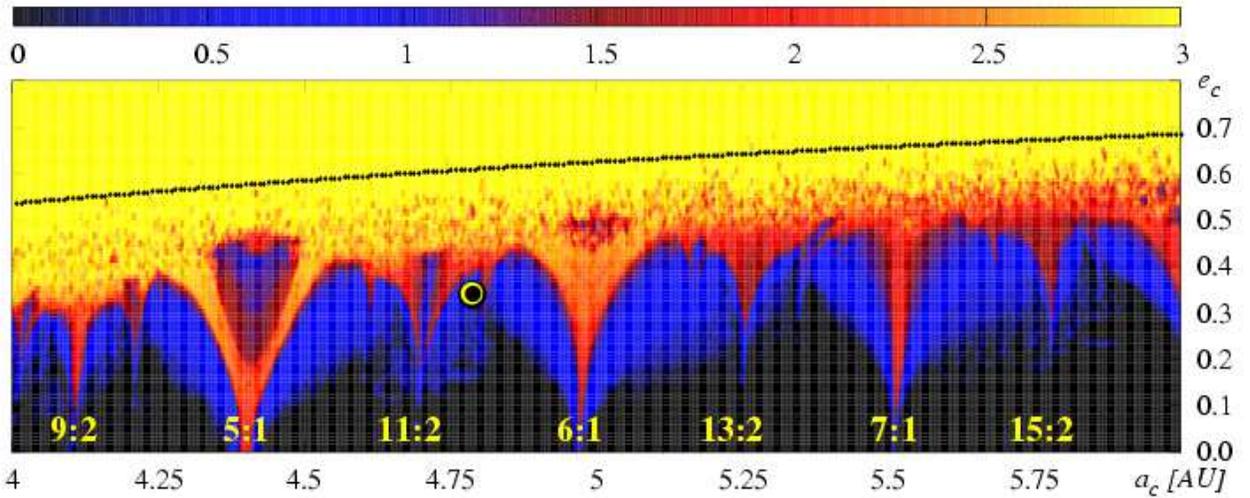}
\caption{
The stability map in the $(a_{\idm{c}},e_{\idm{c}})$-plane, in terms of the
spectral number, $\log SN$, (see the text for an explanation) for the best fit
solution GM1 (see Table~1). Color code is for $\log SN$: black means
quasi-periodic,  regular configurations, and yellow strongly chaotic systems.
Labels are for the identification of the low-order MMRs. A big circle marks the
parameters of the GM1 fit.
}
\label{fig:fig9}
\end{figure}
%
%
\begin{figure}
\centering
\includegraphics[]{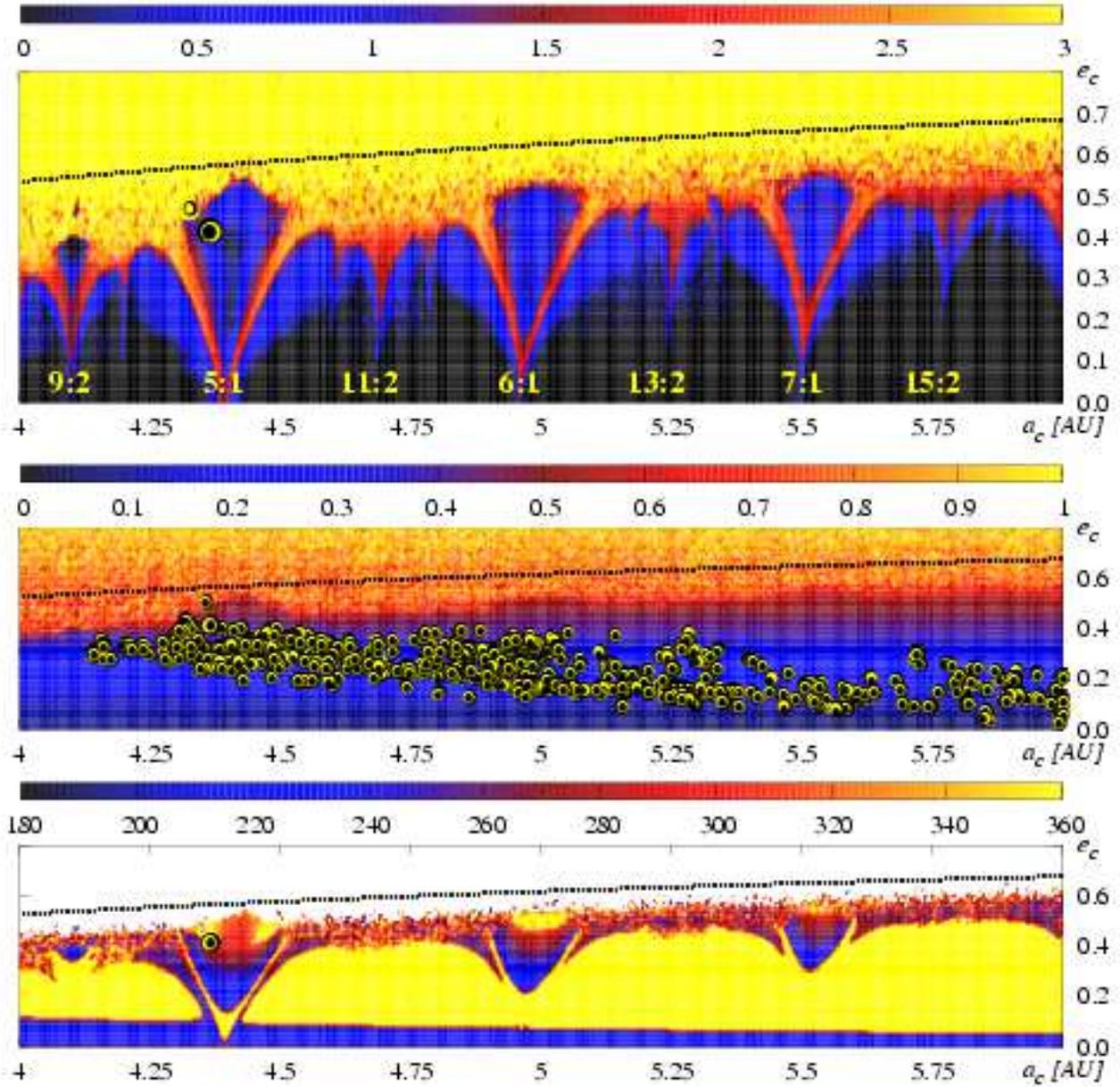}
\caption{
The upper plot is  for the stability map in the
$(a_{\idm{c}},e_{\idm{c}})$-plane, in terms of the spectral number, $\log SN$,
(see the text for an explanation) for the best fit  solution GM2 (see Table~1).
Color code is for $\log SN$: black means quasi-periodic,  regular
configurations and yellow strongly chaotic systems. A big circle marks the
parameters of the GM2 fit. For comparison, the elements of NL1 fit are depicted
as a smaller circle. The middle plot is for maximum eccentricity of the outer
planet attained during the integration period (about $10^5$~yr). Circles in
this plot mark parameters of the fits obtained in the GAMP search within
$1\sigma$ confidence interval of the GM1 fit. The lower plot is for the maximum
of $\theta$. The resolution of the maps is is $600\times100$ data points.
}
\label{fig:fig10}
\end{figure}
%
%
\begin{figure}
\centering
\hbox{
      \includegraphics[]{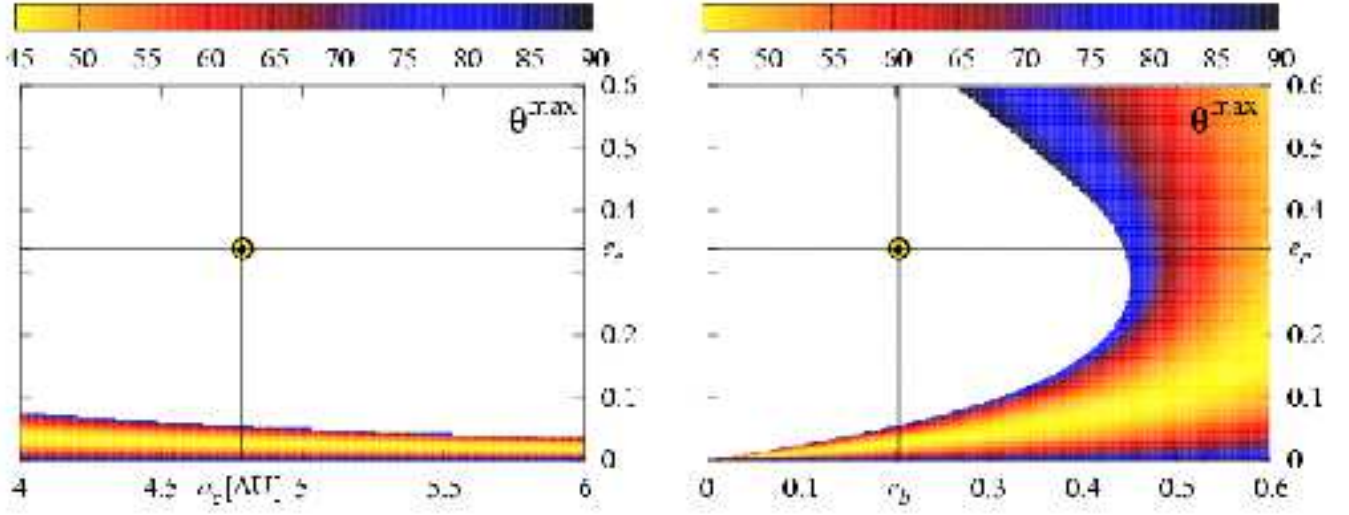}
}
\caption{
The semi-amplitude of the critical argument of SAR, $\theta$, about the
libration center $180^{\circ}$, derived by the application of the secular
octupole theory by \cite{Lee2003,Lee2003a}.  The nominal parameters of the GM1
fit are marked by circles and intersecting lines.  Compare with the lower
plot in Fig.~\ref{fig:fig10}.
}
\label{fig:fig11}
\end{figure}
%
%
\begin{figure}
\centering
\includegraphics[]{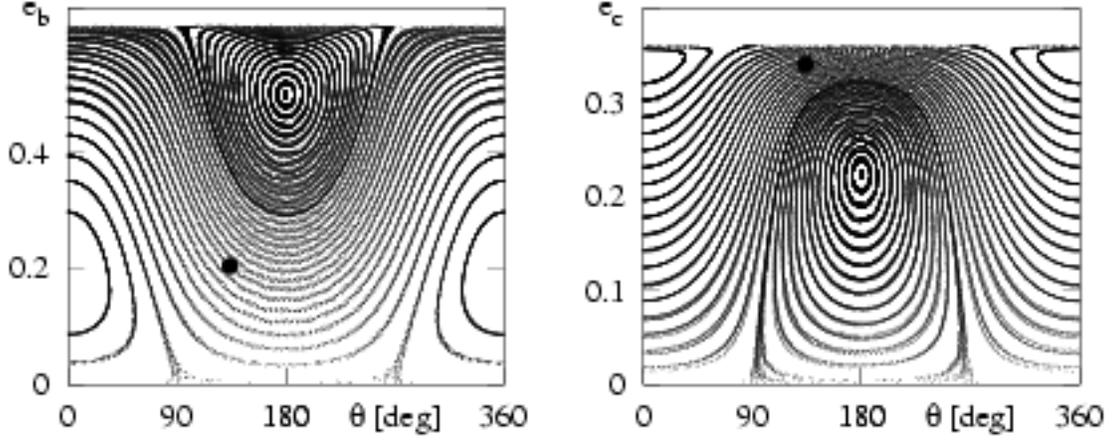}
\caption{
Phase diagram of the GM1 fit obtained by numerical integration of the full,
$N$-body model for constant values of the integrals of the total angular
momentum and total energy. The nominal configuration is marked by a filled
circle  in both panels. See the text for explanation.
}
\label{fig:fig12}
\end{figure}
%
%
\begin{figure}
\centering
\includegraphics[width=17cm]{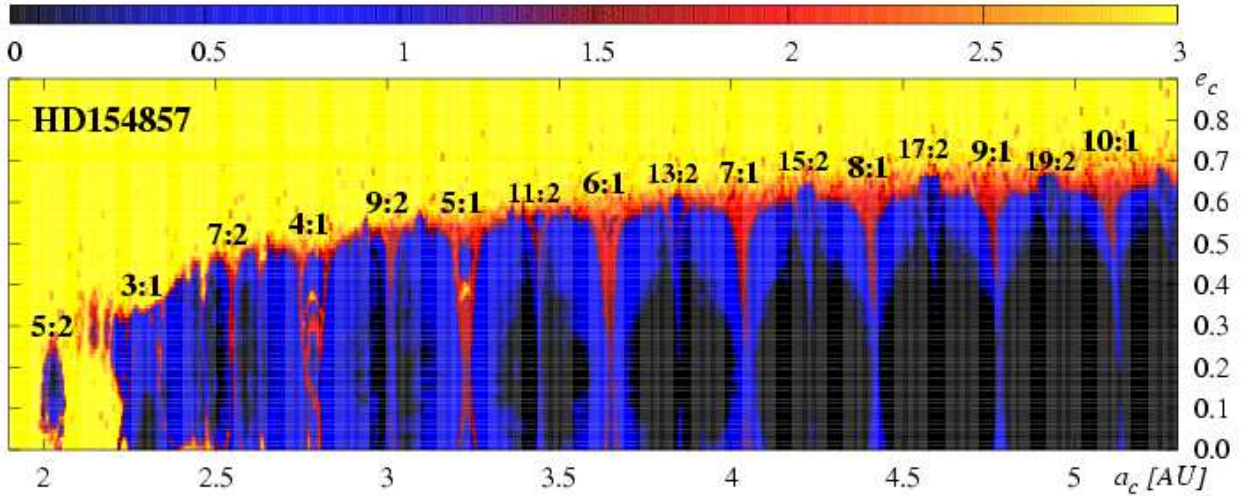}
\caption{
The stability map in the $(a_{\idm{c}},e_{\idm{c}})$-plane, in terms of the
spectral number, $SN$,  for the HD~154857 planetary system. Color code is for $
SN$: black means quasi-periodic,  regular configurations and yellow (light
gray) strongly chaotic systems. Labels are for the identification of the
low-order MMRs. Resolution of the plot is $640\times100$ data points.
}
\label{fig:fig13}
\end{figure}

\end{document}